# Entanglement islands in 1D and 2D lattices with defects

Ivan P. Christov

Physics Department, Sofia University, 1164 Sofia, Bulgaria

**Abstract**

We investigate the spatial structure of quantum entanglement in one- and two-dimensional lattice systems containing structural defects, using the Time-Dependent Quantum Monte Carlo (TDQMC) method. By constructing reduced density matrices from ensembles of guide waves, we resolve spatial variations in both Coulomb-mediated entanglement and coherence without requiring full many-body wavefunctions. This approach reveals localized regions, entanglement islands, where quantum correlations are enhanced or suppressed due to the presence of vacancies or interaction inhomogeneities. In 1D systems, entanglement tends to concentrate near defects, while in 2D we observe bridge-like and radially symmetric domains. Our results demonstrate that TDQMC offers a scalable and physically transparent framework for real-space quantum information analysis, with implications for information transfer in atomic-size structures, quantum materials, entanglement-based sensing, and coherent state engineering.

Keywords: spatial entanglement; lattices; quantum Monte Carlo

## 1. Introduction

Entanglement in complex quantum systems manifests as the inseparability of the many-body wave function - a hallmark of quantum correlations that cannot be explained classically. This entanglement can originate from fundamentally quantum, nonlocal effects, such as the symmetrization of indistinguishable particle wave functions, or from classical interactions like the Coulomb repulsion, which depends explicitly on particle separation [1]. In particular, Coulomb interactions, with their long-range, distance-dependent nature, make it meaningful to ask how entanglement varies spatially across a system. Conventionally, entanglement is quantified using global measures such as von Neumann or Rényi entropies, which compress the entanglement content of an entire system into a single scalar quantity [2]. However, in systems where the interaction landscape is non-uniform, due to geometry, boundaries, or varying interaction strengths, it becomes both natural and necessary to investigate how entanglement distributes in space, giving rise to what we term "entanglement islands".

From a broader perspective, understanding entanglement islands has implications in quantum gravity, where that concept has recently emerged in the study of black holes, where "entanglement islands" refer to regions in spacetime that are included in the entanglement wedge of an observer [3]. While our usage is distinct - focusing on lattice-based many-body systems - the conceptual overlap in both cases is that entanglement islands serve as nontrivial subregions whose entanglement properties are crucial for illuminating structural properties.

Prior work has demonstrated that in a one-dimensional (1D) hydrogen molecule, entanglement is concentrated in the region between the atoms, even before they are bound, emphasizing the physical relevance of where entanglement resides [4]. In two dimensions (2D), entanglement analysis becomes even more intriguing. Here, the notion of an entanglement island gains further dimensional richness, it may refer to regions protected by topological entanglement entropy, or to domains that remain entangled even as the surrounding system undergoes localization, symmetry breaking, or other phase transitions. Moreover, the interplay of geometry and topology in 2D such as periodic boundary conditions or nontrivial lattice symmetries can give rise to entanglement domains that are sharply defined and exhibit long-lived coherence.

Standard simulation techniques often rely on exact diagonalization or Quantum Monte Carlo (QMC), particularly for 1D chains [5], [6]. For 2D systems, where computational costs are significantly higher, entanglement estimators based on Rényi entropies, typically obtained via the replica method within QMC, are commonly employed [7]. Other numerical approaches, such as the density matrix renormalization group (DMRG), face limitations in 2D [8], although recent advances in tensor network methods have enabled more refined investigations of entanglement in low dimensional structures [9] that includes the recent augmented trees method which works well in structured Hilbert spaces [10].

In this work, we explore the structure of entanglement islands in one-dimensional (1D) and two-dimensional (2D) lattice systems including vacancies. By entanglement islands, we refer to localized or semi-localized subregions within a quantum system where entanglement is significantly enhanced or structurally distinct from the surrounding sites. These islands can arise due to lattice potentials, symmetry constraints or topological features, and they offer insight into how quantum information is organized in many-body systems. Here we adopt a numerical framework grounded in the Time-Dependent Quantum Monte Carlo (TDQMC) method [11,12] to study spatial entanglement features in many-body systems. By neglecting nonlocal causality effects, TDQMC circumvents the curse of dimensionality associated with solving the full many-body Schrödinger equation or employing other trajectory-based methods [13], while still capturing essential quantum features such as entanglement and coherence. The method maintains scalability comparable to time-dependent Hartree-Fock, even for larger particle numbers.

The structure of this paper is as follows. Section 2 introduces the models and computational methods, including the Time-Dependent Quantum Monte Carlo formulation, entropy measures, and numerical setups. Section 3 focuses on entanglement islands in 1D and 2D systems, starting from two-particle models and extending to larger chains where for the 2D lattices the role of geometry, defects, and boundary conditions in shaping entanglement is explored. We discuss the results in Section 4.

## 2. Methods

TDQMC recasts the full many-body Schrödinger equation into a self-consistent system of coupled equations for trajectories ("walkers") and associated wave functions ("guide waves"). Each physical particle is described by an ensemble of such walkers in real space and a corresponding ensemble of guide waves. The two evolve together: the walkers respond to changes in the wave field, and the waves evolve according to the positions of the walkers. Using non-local kernels, TDQMC captures the effective interactions between particles across restricted regions of configuration space, in contrast to the exact formulation, where nonlocal causality connects arbitrary points throughout configuration space.

Here we focus on lattice (crystal) systems where the Coulomb interactions between the electrons play a key role in entanglement. When using TDQMC the localized defects are easily obtained through local potential modification, so we don't have to rebuild a huge basis or Hamiltonian to insert a defect. We just remove a potential well or modify the local potential grid, that's cheap, fast, and flexible — unlike in e.g. QMC where careful construction of defect-aware states or trial wave functions would be necessary.

In the general TDQMC framework the quantum state is described by a set of M Schrödinger-like equations for the different guide waves $\varphi_i^k(\mathbf{r}_i,t)$, each one attached to separate trajectory $\mathbf{r}_i^k(t)$, for the *i*-th electron; *i*=1,2,…N, (particle-wave dichotomy, [12]):

$$i\hbar\frac{\partial}{\partial t}\varphi_i^k(\mathbf{r}_i,t)=\left[-\frac{\hbar^2}{2m}\nabla_i^2+V_{en}(\mathbf{r}_i)+V_{eff}^k\left(\mathbf{r}_i,t\right)\right]\varphi_i^k(\mathbf{r}_i,t) \tag{1}$$

with $k=1,\dots,M$ being the number of walkers, $V_{en}(\mathbf{r}_i)$ is the electron-nuclear potential, and where:

$$V_{eff}^k\left(\mathbf{r}_i,t\right)=\sum_{j\neq i}^{N}\frac{1}{Z_{j,i}^k}\sum_{l}^{M}V_{ee}\left[\mathbf{r}_i,\mathbf{r}_j^l(t)\right]K\left[\mathbf{r}_j^l(t),\mathbf{r}_j^k(t),\sigma_j\right] \tag{2}$$

is the effective electron-electron interaction potential represented as a Monte-Carlo (MC) convolution which incorporates the spatial quantum nonlocality (uncertainty) by allowing each walker for a given electron to interact through the electron-electron potential $V_{ee}\left[\mathbf{r}_i,\mathbf{r}_j\right]$ with a group of walkers of other electron within a region determined by the non-local length $\sigma_j\left(\mathbf{r}_j^k\right)$ [11]. For Gaussian kernel we have:

$$K\left[\mathbf{r}_j, \mathbf{r}_j^k(t), \sigma_j\right] = \exp\left(-\frac{\left|\mathbf{r}_j - \mathbf{r}_j^k(t)\right|^2}{2\sigma_j\left(\mathbf{r}_j^k\right)^2}\right), \tag{3}$$

which determines the weighting factor in equation (2) to be:

$$Z_j^k = \sum_{l=1}^{M} K\left[\mathbf{r}_j^l(t), \mathbf{r}_j^k(t), \sigma_j\right] \tag{4}$$

As seen from Equations (1) - (4) the limit $\sigma_j \to \infty$, where $K\left[\mathbf{r}_j, \mathbf{r}_j^k(t), \sigma_j\right] \to 1$, recovers the mean field approximation used in atomic physics [14] while the actual value of $\sigma_{j,i}$ is determined variationally for each concrete system.

The connection between the trajectories $\mathbf{r}_i^k(t)$ and the waves $\varphi_i^k(\mathbf{r}_i, t)$ is given by the walker's velocities, for imaginary-time propagation ($t = -i\tau$):

$$d\mathbf{r}_i^k(\tau) = \mathbf{v}_i^{Dk} d\tau + \boldsymbol{\eta}_i(\tau)\sqrt{\frac{\hbar}{m} d\tau}\,, \tag{5}$$

where the drift velocity is:

$$\mathbf{v}_i^{Dk}(\tau) = \frac{\hbar}{m}\left[\frac{\nabla_i \varphi_i^k(\mathbf{r}_i, \tau)}{\varphi_i^k(\mathbf{r}_i, \tau)}\right]_{\mathbf{r}_i = \mathbf{r}_i^k(\tau)}. \tag{6}$$

Here $\boldsymbol{\eta}(\tau)$ is Markovian stochastic process and the walkers $\mathbf{r}_i^k(\tau)$ sample the moduli square of the corresponding guide waves $\left|\varphi_i^k(\mathbf{r}_i, \tau)\right|^2$.

We examine basic 1D lattices and square 2D lattices, typical for materials such as transition metal oxides (e.g., CuO, NiO), which may contain vacancies. For simplicity, we neglect the effects of spin-orbit coupling and spin-polarized states. Electrons are assumed to be strongly localized near the minima of a soft-core lattice potential, with opposite spins, allowing us to neglect exchange effects to first approximation. Although the electrons are mostly bound to their parent atoms, they experience the long-range Coulomb interaction and can tunnel to neighboring sites and throughout the lattice. Thus, the overall wavefunction for each electron is spread out, similarly to Bloch states in tight-binding approximation (e.g. in [15]).

By treating the ensemble of wavefunctions $\varphi_i^k(\mathbf{r}_i, t)$ generated by TDQMC as random variables, we construct a reduced density matrix for the *i*-th electron. This matrix acts as a variance-covariance matrix in Hilbert space, capturing important statistical information about the quantum state [4], [12]:

$$\rho_i\left(\mathbf{r}_i, \mathbf{r}_i', t\right) = \frac{1}{M}\sum_{k=1}^{M}\varphi_i^{k*}(\mathbf{r}_i, t)\varphi_i^k(\mathbf{r}_i', t) \tag{7}$$

Instead of the more commonly used von Neumann entropy here we quantify the spatial entanglement using linear quantum entropy [15]-[18], for the ground state:

$$S_L^i\left(\rho, t\right) = 1 - Tr\left(\rho_i^2\right) = 1 - \int \rho_i^2\left(\mathbf{r}_i, \mathbf{r}_i\right) d\mathbf{r}_i \tag{8}$$

One advantage of TDQMC is the ability to compute entanglement locally by exploiting the localization of Monte Carlo walkers. A specific region can be selected, and the local quantum entropy computed using the guide waves within that region. This approach forms the basis of a "localized quantum information theory," where the system is treated as many quantum neighborhoods, each with its own entanglement structure:

$$\rho_i^m\left(\mathbf{r}_i, \mathbf{r}_i', t\right) = \frac{1}{M_m}\sum_{k=1}^{M_m}\varphi_i^{k*}(\mathbf{r}_i, t)\varphi_i^k(\mathbf{r}_i', t), \tag{9}$$

where $M_m$ is the number of walkers within region *m,* for the *i*-th electron. Thus, the local entanglement using the local linear entropy is defined as:

$$S_{L,i}^m\left(\rho\right) = \int\left[\rho_i^m\left(\mathbf{r}, \mathbf{r}\right) - \rho_i^{m2}\left(\mathbf{r}, \mathbf{r}\right)\right] d\mathbf{r} \tag{10}$$

where the local density matrices $\rho_i^m(\mathbf{r},\mathbf{r}')$ are to be normalized to unity trace. In this way the TDQMC methodology allows us to calculate entanglement locally without constructing or tracing out the full density matrix - by relying on sampling and local reductions instead.

### 3. Results

We examine lattice potentials $V_{en}(\mathbf{r})$ for 1D and 2D lattices with a lattice constant *d* and a screened Coulomb interaction at each site. Soft-core modification of the Coulomb potential is used to avoid the singularity at the core, which for 2D lattice is provided by:

$$V_{en}(\mathbf{r}) = \sum_{m,n} \frac{V_0}{\sqrt{(x-nd)^2 + (y-md)^2 + a^2}} \exp\left[ -\frac{\sqrt{(x-nd)^2 + (y-md)^2}}{\lambda} \right], \tag{11}$$

which represents a series of localized potentials centered at each lattice site, with the interaction strength controlled by the distance between $\mathbf{r} = (x, y)$ and the site's core. In Eq. (11) (n,m) are the integer coordinates of the lattice sites in the 2D plane, $V_0$ is the Coulomb interaction strength and *a* is the soft-core regularization parameter. The screening effects, characterized by the screening length $\lambda$, describe how the Coulomb interaction between an electron and the lattice is weakened due to the presence of other electrons. Notice that for some strongly correlated systems such as transition metal oxides (e.g., CuO, NiO), that screening can be significant. It is assumed here that the electron-electron repulsion is described by a soft-core Coulomb potential:

$$V_{ee}\left[\mathbf{r}_i, \mathbf{r}_j\right] = \frac{e^2}{\sqrt{\left|\mathbf{r}_i - \mathbf{r}_j\right|^2 + a^2}} \tag{12}$$

As previously noted, the parameters $\sigma_j$ control the fraction of walkers from one electron ensemble that interact with a given walker from another, implementing the spatial nonlocality (or uncertainty) intrinsic to quantum particles. In TDQMC, these parameters are determined by variationally minimizing the system's energy. While one might expect them to scale with the total spatial uncertainty of the whole electron distribution, potentially approaching the lattice size for identical particles, we find the optimized values typically remain within nearest-neighbor distances. This localization is physically justified: unbounded growth of the parameters $\sigma_j$ would suppress electron-electron correlations, asymptotically approaching the Hartree mean-field limit, as pointed out earlier. The calculations were conveniently conducted in atomic units (a.u.) where $e = m = \hbar = 1$ and the parameters in Eq.(11) and Eq.(12) are: $a = 0.4$ , $V_0 = -1$ , $\lambda = 1.11$ .

To access the entanglement entropy locally, we partition the system into 21 zones (strips) along each coordinate axis and apply Eq. (9) to the walkers located within each zone, thereby constructing the reduced density matrix specific to that 1D or 2D region. In all calculations periodic boundary conditions are implemented.

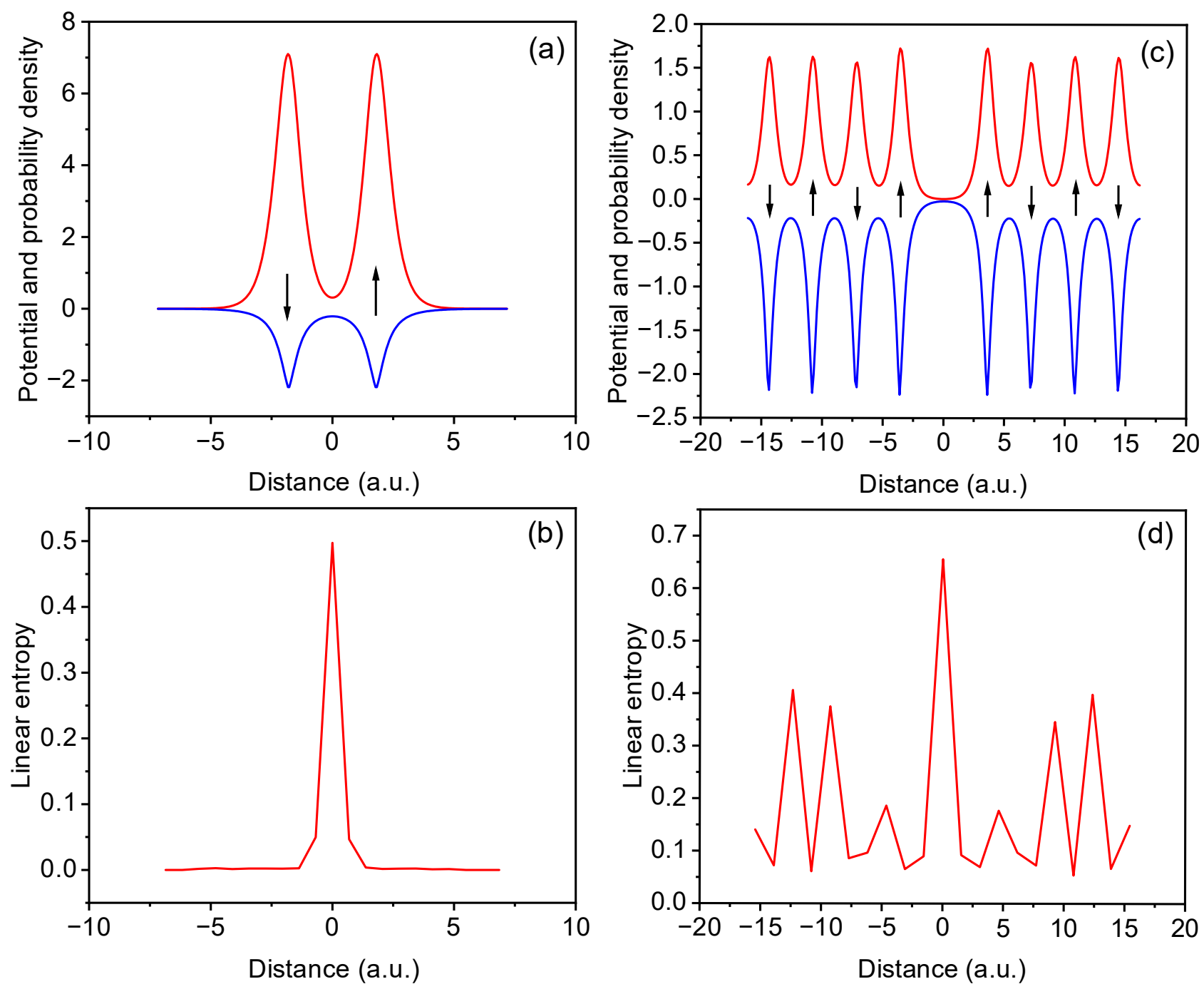

**Figure 1. (a) -** external potential (blue) and electron probability density (red) for a 1D diatomic system. **(b)** - corresponding local entanglement entropy profile, showing maximal entanglement at the midpoint between the two atoms. **(c) -** external potential and probability density for an eight-atom 1D chain with a central vacancy. **(d)** - the associated linear entropy shows enhanced entanglement at the defect site and residual correlations within each sub-chain, highlighting the impact of structural discontinuity on entanglement distribution.

Figures 1a and 1c illustrate the external potentials (blue lines) and spatial distributions of the electrons (red lines) in one-dimensional diatomic and eight-atom lattices, respectively. Although spin degrees of freedom are not explicitly included in this model, the black arrows indicate assumed spin orientations, which serve to justify the neglect of exchange interaction between electrons with identical spin occupying adjacent atomic sites. In the eight-atom system (Fig. 1c), a central vacancy is introduced, effectively breaking the lattice symmetry. The probability density shows spatial separation across the defect. The corresponding linear entropy profiles are shown in Fig. 1b and Fig. 1d. In the diatomic case (Fig. 1b), the linear entropy reaches a maximum in the region between the two atoms, indicating that the entanglement is concentrated where the electron probability densities overlap (also in [4]). Interestingly, our results reveal that the maximum of the local entanglement entropy tends toward the value of the global entanglement entropy. The entropy profile in the eight-atom lattice with a vacancy (Fig. 1d) exhibits a pronounced peak at the defect site, with smaller oscillations aligned with the remaining potential minima, which reflects the localization of quantum correlations at the structural discontinuity. In fact, the central defect in an eight-site chain enhances long-range entanglement across the vacancy while suppressing it in adjacent sites. This indicates that even in systems without explicit spin dynamics, structural discontinuities can effectively sever quantum correlations by disrupting the spatial coherence of the wavefunction.

Extending the analysis to two-dimensional configurations reveals a richer entanglement topology, shaped not only by inter-particle distance but also by lattice symmetry and geometry. Similar to the 1D case, for pair of identical 2D electrons (Fig. 2a–c), the linear entropy is sharply peaked along the region connecting the particles, forming bridge-like structures. These features demonstrate that the Coulomb-mediated entanglement remains spatially localized despite the additional degree of freedom and are spatially confined in zone where the correlated quantum fluctuations of the two atoms are most pronounced. Notice the almost identical results are predicted by the TDQMC method in Fig. 2b and by using exact solution of the two-body Schrodinger equation (Fig. 2c). For more complex 2D geometries with central vacancy (Fig. 2d–e) symmetry-driven patterns which include localized entanglement “islands” are clearly observed. In particular, the entropy maps reveal regions of radial decay and confinement of entanglement near topological discontinuities. These results confirm that entanglement is not only a nonlocal property but is also modulated by spatial geometry and lattice topology, even in systems where spin effects are not explicitly treated.

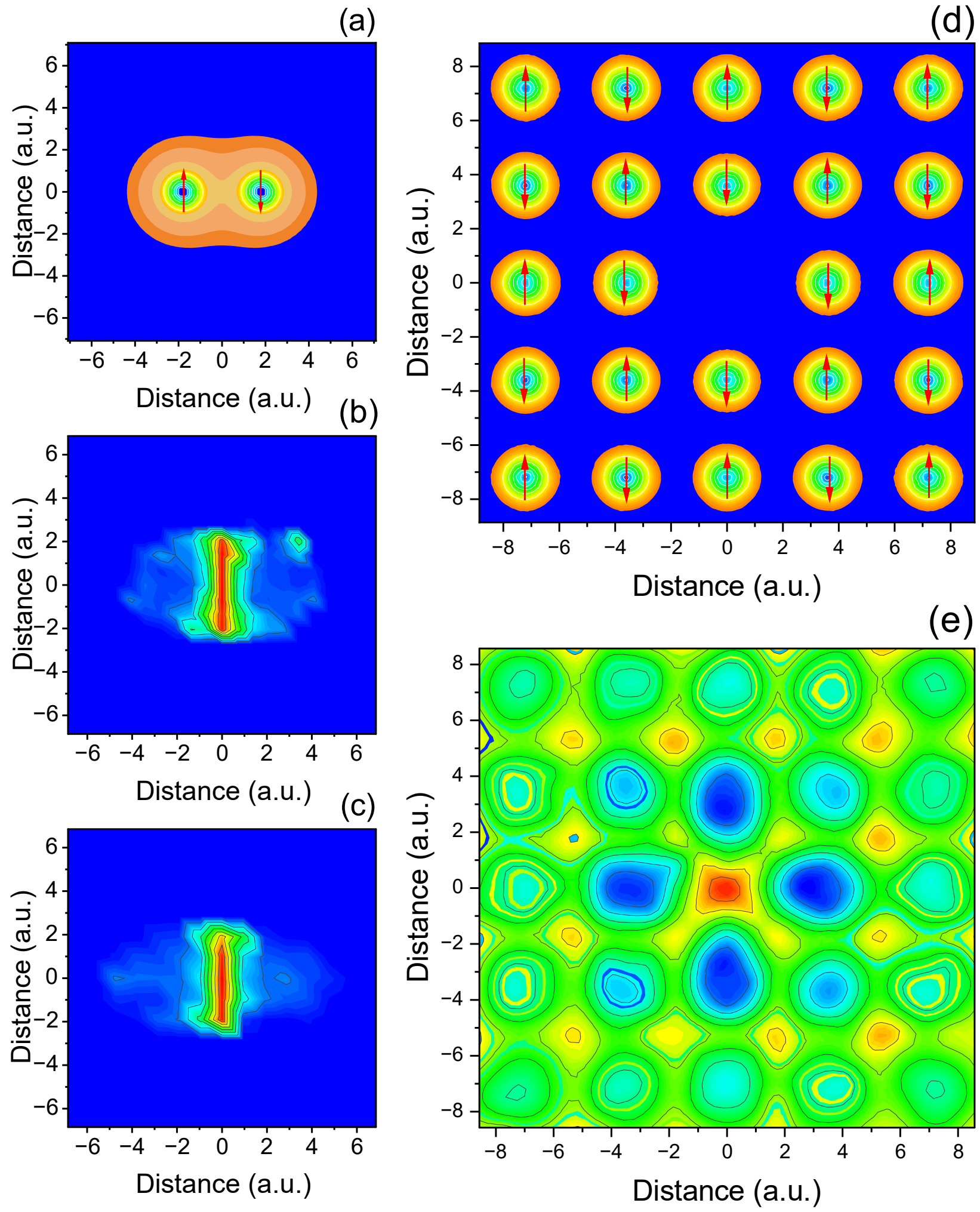

**Figure 2.** Spatial entanglement distribution for two-dimensional diatomic electron configuration - (**a**); TDQMC prediction - (**b**); numerically exact result - (**c**). Entanglement is localized in narrow regions between particles, forming bridge-like structures. **(d–e)** - linear entropy for more complex 2D arrangement, including configuration with a central vacancy. These reveal localized entanglement "islands" and radial enhancement around structural boundaries, demonstrating how geometry modulates quantum correlation topology.

A better understanding of quantum correlations in interacting systems requires analyzing besides entanglement entropy $S_L^m(\rho)$ also the spatial distribution of coherences. While the entanglement entropy reflects how much information is shared across a partition of the system, the relative entropy of coherence $C_r^m(\rho)$ measures the degree of quantum superposition within a subsystem. Plotting these two quantities together reveals whether coherence arises primarily from internal delocalization or from entanglement with other parts of the system. For example, a high coherence with low entanglement indicates a locally coherent, globally separable state, whereas high entanglement with low coherence implies that the local state is highly mixed due to strong correlations. Such analysis provides deeper insight into the distribution of quantum information, distinguishing between local interference effects and true multipartite entanglement, and is particularly useful in systems with spatial structure, disorder, or time-dependent dynamics. In the context of TDQMC, the reduced density matrix of a subsystem, constructed from the calculated guide waves, allows us to observe the interplay between coherence and entanglement, thereby revealing subtle structural transitions or spatial entanglement patterns that might be obscured by global observables.

The relative entropy of coherence measures how much quantum coherence (off-diagonal elements) is "lost" when the system is made purely incoherent (diagonalized). In other words, it quantifies the distance between a density matrix and its decohered (diagonal) version [16]:

$$C_r(\rho) = S(\rho_{diag}) - S(\rho) \tag{13}$$

where $S(\rho) = -Tr[\rho \log \rho]$ is the von Neumann entropy.

While both relative entropy of coherence and von Neumann entropy involve the density matrix and entropy-based measures, they serve distinct purposes: the von Neumann entropy quantifies the degree of mixedness of a quantum state, whereas the relative entropy of coherence measures how much quantum coherence is present relative to a completely incoherent (diagonal) state. In calculations where the electron's wave functions are delocalized and distributed nearly equally across the effective dimensionless area $\Sigma$ of the lattice, the diagonal elements of the reduced density matrix can be approximated as $\rho_{diag} \approx I / \Sigma$, which leads to a nearly constant value of $S_L(\rho_{diag}) \approx 1 - 1/\Sigma$. As a result, the spatial profile of the linear relative entropy of coherence, computed using the linear entropy rather than the von Neumann entropy and shown in Fig. 3, displays an almost mirror-image relationship to the linear entropy distribution shown in Fig. 2e. This inverse correlation highlights the intrinsic trade-off between local coherence and entanglement: regions of low entanglement entropy correspond to nearly pure, locally coherent states, whereas regions of high entanglement reflect mixed states arising from strong correlations and a corresponding suppression of coherence. It should be noted, however, that despite the simple interpretation offered by the linear entropy of coherence, it can serve as an objective measure of coherence only in relatively simple systems, such as the 1D and 2D diatomics shown in Fig. 1a and Fig. 2a. For more complex structures with defects, the linear entropy of coherence may not reflect the true relative entropy of coherence of Eq. 13. This is demonstrated in Fig. 3b, where the region of reduced coherence extends to larger distances from the defect than those predicted by the linear entropy of coherence in Fig. 3a.

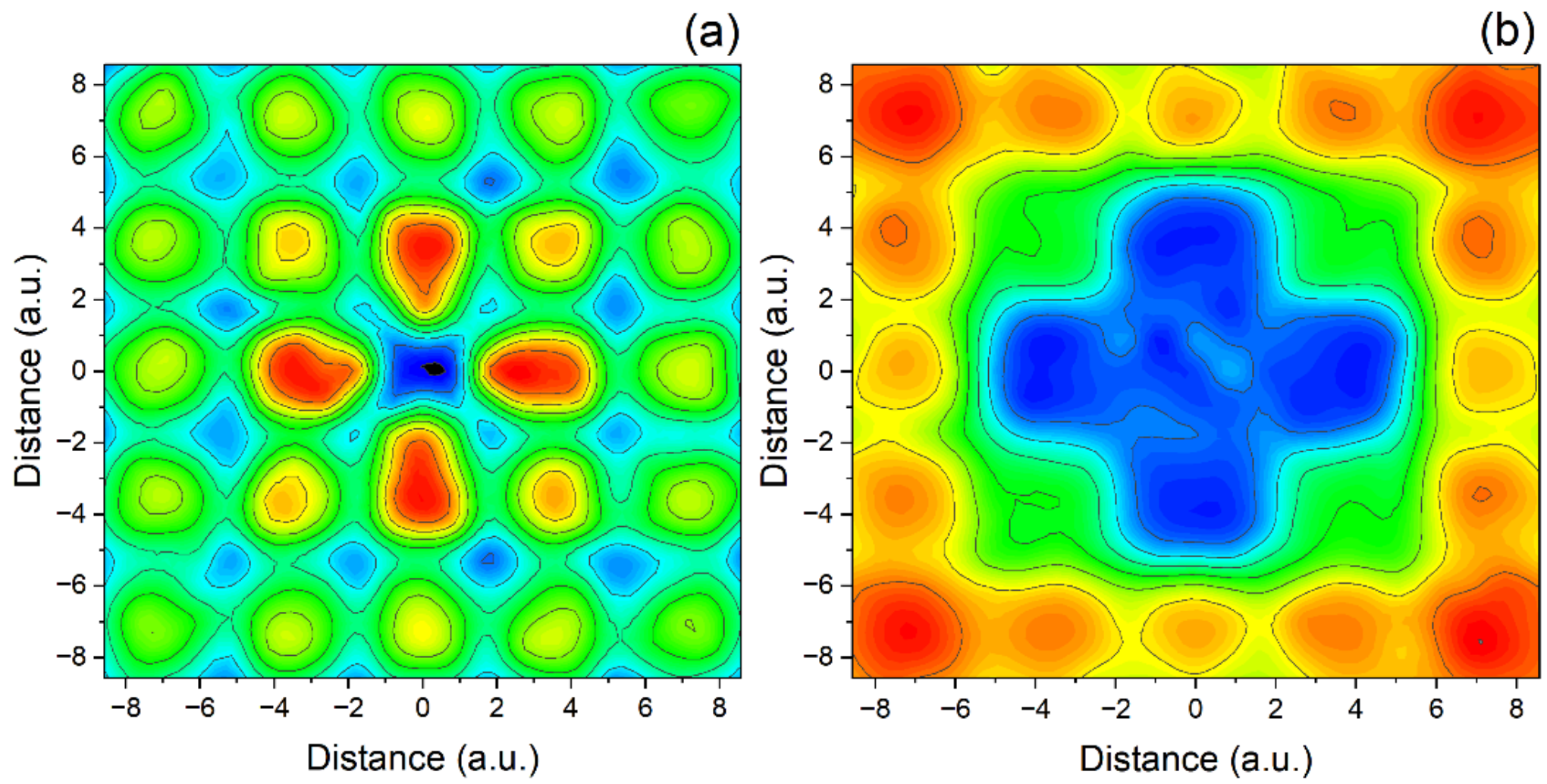

**Figure 3.** Spatial distribution of the linear entropy of coherence $C_L(\rho)$ - **(a)**, and the relative entropy of coherence $C_r(\rho)$ - **(b)**, computed from the reduced density matrix obtained via TDQMC. Regions of high coherence (red) correspond to strong local superposition in the position basis, while low coherence (blue) reflects increased mixedness due to entanglement with the other particles. The profile in (a) exhibits an inverse relationship to the entanglement entropy in Fig. 2e, consistent with the trade-off between local coherence and multipartite entanglement.

### 4. Conclusions

Our findings indicate that entanglement islands are robust features in systems with spatially modulated interactions or boundary-induced inhomogeneities. In particular, we observe that in 1D hydrogen-like models, the entanglement entropy is highest in the interatomic region and persists even when the atoms are far apart, revealing the non-local character of quantum correlation. In extended lattice systems, we find that entanglement depends strongly on particle density modulations and interaction hotspots, suggesting a deep link between entanglement localization and emergent structure in many-body states. Although for convenience we used the linear entropy to quantify entanglement, it may sometimes fail to reproduce the behavior of the von Neumann entropy in low-density systems as shown recently [21]. However, repeating the key calculations confirmed that, for the systems and parameter ranges considered here, the spatial profile of the entanglement obtained from the von Neumann entropy remains qualitatively unchanged. This behavior does not persist when the relative entropy of coherence is computed using the von Neumann entropy instead of the linear entropy. In that case, the region of reduced coherence surrounding the defect can be substantially broader.

The TDQMC method we use is built to make it easy to trace out variables, because the wavefunctions are sampled with correlations - not approximated as independent orbitals. Unlike Density Functional Theory or QMC which often rely on marginal densities or don't represent the full many-body wavefunction, TDQMC retains a correlated description of the wavefunction via ensembles of walkers and their guide wavefunctions. This allows for the faithful computation of reduced density matrices and spatially resolved entanglement measures, not merely observables derived from them. In this way, TDQMC provides a microscopic, entanglement-aware simulation framework for crystalline and interacting systems, enabling the visualization and quantification of quantum correlations with high spatial resolution.

These insights can be extended beyond lattice-bound electrons or model bosonic systems. The TDQMC framework is readily adaptable to other controllable quantum platforms such as trapped ions or ultracold atoms in optical lattices, where interparticle interactions and spatial arrangements can be precisely engineered. In such systems, spatially resolved entanglement structures, such as the entanglement islands observed here, may play a critical role in understanding entanglement propagation, localization phenomena, or the stability of quantum phases under external modulation.

From a broader perspective, the local spatial entanglement structures have implications not just for condensed matter and quantum information, but also for quantum chemistry and field theory. While our usage is rooted in lattice quantum systems, the underlying theme is similar: understanding how quantum information is partitioned across space, and which regions are included in a subsystem's entanglement, is essential for interpreting global properties of a system.

**Acknowledgement**

This research is based upon work supported by the Bulgarian Ministry of Education and Science as a part of National Roadmap for Research Infrastructure, grant D01-351/13.12.2023 (ELI ERIC BG), and by the National Science Fund , grant КП-06-Н78/6.